\documentstyle[12pt]{article}
\newcommand{\beq}{\begin{equation}}
\newcommand{\eeq}{\end{equation}}
\newcommand{\beqa}{\begin{eqnarray}}
\newcommand{\eeqa}{\end{eqnarray}}
\newcommand{\ba}{\begin{array}}
\newcommand{\ea}{\end{array}}

\begin{document}

\begin{flushright}
Preprint CAMTP/97-4\\
August 1997\\
\end{flushright}

\begin{center}
\large
{\bf WKB corrections to the energy splitting \\
in double--well potentials} \\
\vspace{0.25in}
\normalsize
Marko Robnik$^{(*)}$\footnote{e--mail: robnik@uni-mb.si} and 
Luca Salasnich$^{(+)}$\footnote{e--mail: salasnich@math.unipd.it} \\
\vspace{0.2in}
$^{(*)}$ Center for Applied Mathematics and Theoretical Physics,\\
University of Maribor, Krekova 2, SLO--2000 Maribor, Slovenia\\
\vspace{0.2in}
$^{(+)}$ Dipartimento di Matematica Pura ed Applicata \\
Universit\`a di Padova, Via Belzoni 7, I--35131 Padova, Italy \\
Istituto Nazionale di Fisica Nucleare, Sezione di Padova,\\
Via Marzolo 8, I--35131 Padova, Italy \\
Istituto Nazionale di Fisica della Materia, Unit\`a di Milano, \\
Via Celoria 16, I--20133 Milano, Italy
\end{center}

\vspace{0.3in}

\normalsize

\noindent
{\bf Abstract.} By using the WKB quantization 
we deduce an analytical formula 
for the energy splitting in a double--well potential 
which is the usual Landau formula 
with additional quantum corrections. 
Then we analyze the accuracy of our formula for 
the double square well potential 
and the parabolic double--well potential. 

\vspace{0.4in}

PACS numbers: 03.65.-w, 03.65.Ge, 03.65.Sq 
\normalsize
\vspace{0.1in}
  
\newpage

\par
In this paper we analyze the energy splitting 
in a generic one--dimensional double--well potential. 
By using the WKB quantization we deduce an analytical formula 
for the energy splitting which is the usual Landau (1973) formula 
with additional quantum corrections. 
The formula is based on a linear approximation of the 
potential near the turning points. We study the 
validity of this formula for the double square well potential 
and then we discuss the case of a parabolic double--well potential. 
\par
Let us consider a one--dimensional system with Hamiltonian
\beq
H = {p^2\over 2m} + V(x) \; ,
\eeq 
where $V(-x)=V(x)$ is a symmetric double--well potential. 
The stationary Schr\"odinger equation of the system reads 
\beq
{\hat H} \psi (x) = \big( -{\hbar^2\over 2m} {d^2 \over dx^2} + V(x) \big) 
\psi (x) = E \psi (x) . 
\eeq
The Sturm--Liouville theorem (see, for example, Fl\"ugge, 1971) ensures 
that for one--dimensional systems there are no degeneracies in the 
spectrum. Let $\psi_1$ and $\psi_2$ be 
two eigenfunctions of the Schr\"odinger equation 
\beq
{\hat H} \psi_1 = E_1 \psi_1 \;\;\;\;\; \hbox{and} \;\;\;\;\; 
{\hat H} \psi_2 = E_2 \psi_2 \; ,
\eeq
such that $\psi_1(-x)=\psi_1(x)$ and $\psi_2(-x)=-\psi_2(x)$ 
and $E_1\simeq E_2$. To calculate the splitting 
$\Delta E = E_2 - E_1$, we multiply the first equation by $\psi_2$ 
and the second by $\psi_1$ and then we subtract the two resulting 
equations. By integrating from $0$ to $\infty$ we find
\beq
\Delta E = {\hbar^2 \over 2 m} 
{ \psi_1(0)\psi{'}_2(0) - \psi{'}_1(0)\psi_2(0) \over 
\int_0^{\infty} \psi_1(x) \psi_2(x) dx } \; .
\eeq
We write the eigenfunctions $\psi_1$ and $\psi_2$ in terms 
of the right--localized function 
\beq
\psi_0(x) = {1\over \sqrt{2}}(\psi_1(x) + \psi_2(x) )\; .  
\eeq
It is easy to show that $E_0 = <\psi_0 | {\hat H} |\psi_0 > 
= {1\over 2} (E_1 + E_2)$. Then, with the approximation
\beq 
\int_0^{\infty} \psi_1(x) \psi_2(x) dx \simeq {1\over 2} 
\int_0^{\infty} \psi_0^2(x) dx = {1\over 2} \; ,
\eeq
we get 
\beq
\Delta E = {2\hbar^2 \over m} \psi_0(0)\psi{'}_0(0) \; ,
\eeq
which is the starting formula to calculate the energy splitting. 
One should observe that this quantity is always positive,
because the tail of the right localized eigenfunction
$\psi_0(x)$ at $x=0$ has the same sign for $\psi_0(0)$
and its derivative $\psi'_0(0)$. Another way to see this is
to realize that because of Sturm-Liouville theorem there
are no degeneracies and that implies that all pairs of
almost degenerate states, from the ground state up,
are grouped by odd state above the even state.
\par
To determine the function $\psi_0$ we perform a WKB expansion 
of the Schr\"odinger equation. We observe that 
a generic eigenfunction $\psi$ of the Schr\"odinger equation 
can always be written as
\beq
\psi (x) = \exp{ \big( {i\over \hbar} \sigma (x) \big) } \; ,
\eeq
where the phase $\sigma (x)$ is a complex function that satisfies 
the differential equation
\beq
\sigma{'}^2(x) + ({\hbar \over i}) \sigma{''}(x) = 2m(E - V(x)) \; .
\eeq
The WKB expansion for the phase is given by
\beq
\sigma (x) = \sum_{k=0}^{\infty} ({\hbar \over i})^k \sigma_k(x) \; .
\eeq
Substituting (10) into (9) and comparing like powers of $\hbar$ gives 
the recursion relation ($n>0$) (see Bender, Olaussen and Wang, 1977) 
\beq
\sigma{'}_0^2=2m(E-V(x)), \;\;\;\; 
\sum_{k=0}^{n} \sigma{'}_k\sigma{'}_{n-k}
+ \sigma{''}_{n-1}= 0 \; .
\eeq
With the momentum $p=\sqrt{2m(E-V(x))}$ the first three orders in the 
WKB expansion give 
\beq
\sigma{'}_0 = p\; , \;\;\;\;\; 
\sigma{'}_1 = -{p{'}\over 2p} \; ,
\;\;\;\;\;
\sigma{'}_2 = {p{''}\over 4 p^2} -{3\over 8}{p{'}^2\over p^3} \; ,
\eeq
from which we obtain
\beq
\sigma_0 = \int p dx \; ,
\;\;\;\;\;
\sigma_1 = -{1\over 2} \ln{p} \; ,
\;\;\;\;\;
\sigma_2 = {p{'}\over 4p^2} + {1\over 8} \int {p{'}^2\over p^3 } dx \; .
\eeq
It follows that, up to the second order, the wave--function can be written as 
\beq
\psi (x) = {1\over \sqrt{p}}
\exp{\Big( {i\over \hbar} \int p dx 
- i \hbar \big[{p{'}\over 4p^2} + {1\over 8} 
\int {p{'}^2\over p^3 } dx \big] \Big) } \; .
\eeq
In particular, if we call $a$ and $b$ the two turning points 
corresponding to the energy $E$, 
the right localized wave--function $\psi_0$ is given by
\beq
\psi_0(0) = {C \over \sqrt{|p|}} 
\exp{\Big( -{1\over \hbar}\int_x^a |p| dx 
+ \hbar \big[ {1\over 4}|{p{'}\over p^2}| 
+ {1\over 8} \int_x^a |{p{'}^2\over p^3 }| dx \big] \Big) } \; ,
\eeq
for $0<x<a$ (forbidden region), and 
$$
\psi_0(0) = {C_1 \over \sqrt{p}} 
\exp{ \Big( {i\over \hbar}\int_a^x p dx 
-i\hbar \big[{p{'}\over 4p^2} + {1\over 8} \int_a^x {p{'}^2\over p^3 } dx 
\big] \Big) } + 
$$
\beq
+ {C_2 \over \sqrt{p}} 
\exp{\Big( -{i\over \hbar}\int_a^x p dx 
+ i \hbar \big[{p{'}\over 4p^2} + {1\over 8} \int_a^x {p{'}^2\over p^3 } dx 
\big] \Big) } \; , 
\eeq
for $a<x<b$ (allowed region). 
\par
It is possible to write $C_1$ and $C_2$ in terms of $C$ by imposing 
the uniqueness of the wave--function $\psi_0$ at the turning points. 
Following Landau (1973) we suppose that near the turning point $x=a$ 
it is possible to approximate the potential locally linearly by writing 
\beq
E - V(x) = F_0 (x-a) \; ,
\eeq
with $F_0 > 0$. In this case the connections at the turning point 
imply that 
\beq
C_1 = C e^{i {\pi \over 4}}  \;\;\;\;\; \hbox{and} \;\;\;\;\; 
C_2 = C e^{-i {\pi \over 4}} \; ,
\eeq
and the right localized function $\psi_0(x)$ can be written 
for $a<x<b$ as 
\beq
\psi_0(x) = {2 C \over \sqrt{p}} \cos{\Big( {1\over \hbar}\int_a^x p dx 
-\hbar \big[{p{'}\over 4p^2} + {1\over 8} \int_a^x {p{'}^2\over p^3 } dx 
\big] +{\pi \over 4} \Big) } \; . 
\eeq
To determine $C$ we impose the normalization condition 
\beq
1 = \int_{0}^{\infty} |\psi_0(x)|^2 dx \; ,
\eeq
from which we get 
\beq
C = {1\over 2} [\int_a^b {dx\over p} ]^{-1/2} \; .
\eeq
Here we have replaced the oscillatory $\cos^2(.)$ factor in the
integrand by its average value $1/2$, because of the rapid 
oscillations at small value of $\hbar$. 
The final formula is given by 
\beq
\Delta E = \Delta E_{Landau} \cdot \Delta E_{WKB} \; ,
\eeq
where 
\beq
\Delta E_{Landau} = 
{2\hbar C^2\over m} \exp{\Big( -{1\over \hbar}\int_{-a}^a |p| dx \Big) } 
\eeq
is the usual Landau (1973) formula for the energy splitting and  
\beq
\Delta E_{WKB} = 
(1+{\hbar^2 \over 4} |{p{''}(0)\over p(0)^3}| ) 
\exp{\Big( {\hbar \over 8}\int_{-a}^a |{p{'}^2\over p^3}| dx \Big) } \; 
\eeq 
is the first quantum correction. We note that 
higher--order WKB corrections quickly increase in complexity 
(Robnik and Salasnich, 1997) but, in principle, 
they can be calculated from the equation (11). 
It is important to stress that 
our formula is good if the potential is sufficiently smooth 
so that the linear approximation is valid near the turning points. 
\par
As an example, we consider the double square well potential. 
In this case the linear approximation of the potential near the turning 
point is not valid. The potential is given by  
\beq
\ba{cc}
V(x) = V_0    & \hbox{ for } |x| < a  \\
V(x) = 0      & \hbox{ for } a < x < b \\
V(x) = \infty & \hbox{ for } |x| > b 
\ea
\eeq
For this potential we have $p{'}(x)=p{''}(x)=0$ for $-a<x<a$ and 
the corrections to the Landau (1973) formula are zero. 
A naive application of the splitting formula gives 
\beq
\Delta E = {2\hbar \sqrt{E} \over \sqrt{2m} (b-a)} 
\exp{\Big(- {2a\over \hbar} \sqrt{2m(V_0-E)} \Big)} \; .
\eeq 
This formula is not correct. In fact, by using the exact 
wave--function
\beq
\psi_0(0) = D \exp{\Big(-{1\over \hbar}\sqrt{2m(V_0-E)} \Big)} \; ,
\eeq
for $0<x<a$ (forbidden region), and 
\beq
\psi_0(0) = A \exp{\Big({i\over \hbar}\sqrt{2m(E-V_0)} \Big) } + 
B \exp{\Big(-{i\over \hbar}\sqrt{2m(E-V_0)} \Big) } \; , 
\eeq
for $a<x<b$ (allowed region), and by imposing the exact matching 
and normalization conditions (Fl\"ugge, 1971) we find
\beq
A = {D\over 2} \big( 1 - i \sqrt{V_0-E\over E} \big) 
\exp{\Big( {a\over \hbar} \sqrt{2m(V_0-E)} - {a\over \hbar}\sqrt{E} \Big)} \; ,
\eeq
\beq
B = {D\over 2} \big( 1 + i \sqrt{V_0-E\over E} \big) 
\exp{\Big( {a\over \hbar} \sqrt{2m(V_0-E)} + {a\over \hbar}\sqrt{E} \Big)} \; ,
\eeq
and 
\beq
D^2 = {2E\over V_0 (b-a)} 
\exp{\Big( - {2a\over \hbar} \sqrt{2m(V_0-E)} \Big)} \; .
\eeq
Then, by applying the equation (7), we obtain: 
\beq
\Delta E = {4 \hbar E \sqrt{2m(V_0-E)} \over m V_0 (b-a)} 
\exp{\Big( -{2a\over \hbar} \sqrt{2m (V_0 -E)} \Big) } \; .
\eeq
This is the exact energy splitting for the double square well potential. 
It differs by a factor $4\sqrt{E(V_0-E)}/V_0$ from the WKB result 
based on the connection formulae (18) which are not justified 
in the present case. 
\par
To conclude we study a potential where our WKB splitting formula 
can be applied. We consider the parabolic double--well 
potential given by
\beq
\ba{cc}
V(x) = V_0 - x^2 & \hbox{ for } |x| < \sqrt{V_0}  \\
V(x) = 0         & \hbox{ for } \sqrt{V_0} < x < b   \\
V(x) = \infty    & \hbox{ for } |x| > b \\
\ea
\eeq
The turning points are at $x=\pm a$, where $a=\sqrt{V_0-E}$. We observe that 
\beq
\int_a^b {dx\over p} = \int_{\sqrt{V_0-E}}^{\sqrt{V_0}} 
{1\over \sqrt{2m (E-V_0+x^2)} } dx + 
\int_{\sqrt{V_0}}^b {1\over \sqrt{2mE}} dx  \; ,
\eeq
and 
\beq
|{p{''}(0)\over p(0)^3}| = {1\over 2m (V_0-E)^2} \; .
\eeq 
It follows that the energy splitting up to the second order in 
the WKB corrections of $\hbar$ reads  
\beq
\Delta E = {2\hbar C^2\over m} (1+{\hbar^2 \over 8 m (V_0-E)^2}) 
\exp{\Big( -{m\pi \over \hbar} (V_0-E) \Big) } \; , 
\eeq
where 
\beq
C^2 = {\sqrt{2m}\over 2} 
\Big[\ln{ \Big( { \sqrt{E}+\sqrt{V_0}\over\sqrt{V_0-E} } \Big) }  
+ {(b-\sqrt{V_0})\over \sqrt{E} }  \Big]^{-1} \; .  
\eeq

\vspace{0.3in}

\noindent
In this work we have taken the first step towards a systematic
improvement of the Landau formula (Landau and Lifshitz 1973),
which is the semiclassical leading order energy level splitting
formula for pairs of almost degenerate levels in double well
potentials. We have calculated explicitly the quantum corrections
up to the second order, given in equations (21-24).
Our approach is based on the usual WKB expansion
in 1-dim potentials, embodied in formulae (8-11). Thus the
calculation of higher corrections can in principle be continued by
the same method, although the structure of higher terms
increases in complexity very quickly. We have also shown
what happens in cases where the assumption implicit in the
Landau formula (namely the linearity of the potential around
the turning points) is not satisfied: We get a different
result even in the leading semiclassical order, and this
has been shown for the double square well  potential
in equations (26) and (32).

It is our goal to work out
a more direct WKB approach to the solution of the same
problem, by the contour integration technique, based
on requiring the single valuedness of the eigenfunction,
as has been done in the case of a single well potential
in (Robnik and Salasnich 1997). This is our future project.

\section*{Acknowledgments}
This work was supported by the Ministry of Science and
Technology of the Republic of Slovenia, and by the Rector's
Fund of the University of Maribor.

\newpage

\section*{References} 
\parindent=0. pt

Bender C M, Olaussen K and Wang P S 1977 {\it Phys. Rev. } D {\bf 16} 1740 
\\\\
Fl\"ugge S 1971 {\it Practical Quantum Mechanics I} (Berlin: Springer)
\\\\
Gutzwiller M C 1990 {\it Chaos in Classical and Quantum Mechanics} 
(New York: Springer) 
\\\\
Landau L D and Lifshitz E M 1973 {\it Nonrelativistic Quantum Mechanics} 
(Moscow: Nauka) 
\\\\
Maslov V P and Fedoriuk M V 1981 {\it Semi-Classical Approximations in 
Quantum Mechanics} (Boston: Reidel Publishing Company)
\\\\
Robnik M and Salasnich L 1997 {\it J Phys A: Math. Gen.} {\bf 30} 1711 
\\\\
Robnik M and Salasnich L 1997 {\it J Phys A: Math. Gen.} 
{\bf 30} 1719 
\\\\
Voros A 1983 {\it Ann. Inst. H. Poincar\'e} A {\bf 39} 211 

\end{document}